\documentclass[aps,pra,epsf,superscriptaddress,amsmath,amssymb,amsfonts,twocolumn,showpacs,nofootinbib]{revtex4-1}

\usepackage{graphicx}
\usepackage{dcolumn}
\usepackage{bm}
\usepackage{braket}
\usepackage{amsmath}
\usepackage{mathtools}
\usepackage{graphicx,color,xcolor}
\usepackage{hyperref}
\usepackage[capitalize]{cleveref}
\newcommand{\abs}[1]{\left| #1 \right|} 
\usepackage{multirow}
\DeclareMathOperator{\sech}{sech} 

\usepackage[normalem]{ulem} 

\begin{document}
\title{Stability and dynamics of nonlinear excitations\\ in a two-dimensional droplet-bearing environment}

\author{G. Bougas}
\affiliation{Department of Physics, Missouri University of Science and Technology, Rolla, MO 65409, USA}

\author{G. C. Katsimiga}
\affiliation{Department of Physics, Missouri University of Science and Technology, Rolla, MO 65409, USA}%

\author{P. G. Kevrekidis}
\affiliation{Department of Mathematics and Statistics, University of Massachusetts Amherst, Amherst, MA 01003-4515, USA}%

\author{S. I. Mistakidis}
\affiliation{Department of Physics, Missouri University of Science and Technology, Rolla, MO 65409, USA}

\date{\today}

\begin{abstract} 

We unravel stationary states in the form of dark soliton stripes, bubbles, and kinks embedded in a two-dimensional droplet-bearing setting emulated by an extended Gross-Pitaevskii approach. 
The existence of these configurations is corroborated through an effectively reduced potential picture demonstrating their concrete parametric regions of existence. 
The excitation spectra of such configurations are analyzed within the Bogoliubov-de-Gennes framework exposing the destabilization of dark soliton stripes and bubbles, while confirming the stability of droplets, and importantly unveiling spectral stability of the kink against transverse excitations. 
Additionally, a variational approach is constructed providing access to the transverse stability analysis of the dark soliton stripe for arbitrary chemical potentials and widths of the structure. 
This is subsequently compared with the stability analysis outcome demonstrating very good agreement at small wavenumbers. 
Dynamical destabilization of dark soliton stripes via the snake instability is showcased, while bubbles are found to feature 
both a splitting into a gray soliton pair and a transverse
instability thereof. 
These results shed light on unexplored stability and instability properties of nonlinear excitations in environments featuring a competition of mean-field repulsion and beyond-mean-field
attraction that can be probed by state-of-the-art experiments. 

\end{abstract}

\maketitle

\section{Introduction}

Quantum droplets have been recently experimentally realized in single-\cite{bottcher2020new,chomaz2022dipolar} and two-component dipolar gases~\cite{trautmann2018dipolar,politi2022interspecies} and also in homonuclear~\cite{cheiney2018bright,semeghini2018self,cabrera2018quantum} and heteronuclear~\cite{d2019observation,burchianti2020dual} short-range interacting bosonic mixtures. 
Their ultradilute liquid nature manifests by their ability to assemble into flat-top density distributions~\cite{astrakharchik2018dynamics}, their incompressible character~\cite{ferioli2019collisions} and surface tension~\cite{ancilotto2018self}. 
Droplets may arise in completely different contexts such as liquid
Helium~\cite{barranco2006helium}, and photonic 
systems~\cite{michinel1,wilson2018observation,mivehvar2021cavity}. This fascinating phase of matter heralded a new era of investigations devoted to probing quantum fluctuation phenomena in many-body weakly interacting cold atom  settings~\cite{malomed2020family,luo2021new,mistakidis2023few}. 
Quantum fluctuations are strongly dimension-dependent~\cite{zin2018quantum,ilg2018dimensional} and are commonly modeled by the perturbative Lee-Huang-Yang energy (LHY) correction term~\cite{lee1957eigenvalues}, see e.g. also Refs.~\cite{ota2020beyond,mistakidis2021formation,parisi2019liquid} for residual beyond LHY effects. 
The latter is proved to be imperative for droplet formation
as it is its competition with the mean-field (MF) effects
that gives rise to such coherent structures in the
realm of a suitably extended Gross-Pitaevskii equation (eGPE)~\cite{petrov2016ultradilute,petrov2015quantum}. Within this framework a plethora of droplet properties has been described such as their collisional dynamics~\cite{ferioli2019collisions,hu2022collisional,astrakharchik2018dynamics}, collective excitations~\cite{fei2024collective,tylutki2020collective,englezos2023correlated}, the conditions under which modulational instability occurs~\cite{mithun2020modulational,mithun2021statistical,otajonov2022modulational} as well as their characteristic thermal effects~\cite{de2021thermal,boudjemaa2021many}.

Another interesting, yet less explored research direction refers to the study of nonlinear excitations building upon a droplet background, as well as the existence of other self-bound solutions forming in such environments and, in particular,
in the realm of the above mentioned eGPE model featuring the
intriguing competition of MF and LHY-associated nonlinearities. 
For instance, recently it was demonstrated that one-dimensional (1D) droplets can support stable dark soliton configurations while trains of such composite objects are generically unstable~\cite{saqlain2023dragging,katsimiga2023solitary,kopycinski2023ultrawide,edmonds2023dark}. Moreover, the stability of two-dimensional (2D) vortex-droplet states of different charges~\cite{li2018two} but also the dynamical generation of vortices through droplet rotation~\cite{tengstrand2019rotating,yougurt2023vortex,cheng2023dynamics,nikolaou2023rotating}, as well
as defect dragging through the condensate~\cite{saqlain2023dragging} were previously reported. 
A characteristic mechanism of vortex production appearing in repulsive Bose gases~\cite{Anderson_watching_2001,dutton2001observation}, but also in nonlinear optics~~\cite{mamaev1996propagation}, is the well known “snake” (transverse) instability that dark soliton stripes (DSS) undergo once exposed to a higher dimensional geometry such as the 2D one considered herein. 
One of the principal aims of the present work is to explore
the potential of the eGPE model to feature such an instability.
This phenomenon is indeed well-documented~\cite{feder2000dark,muryshev2002dynamics,mateo2014chladni,katsimiga2017many} as well as the  conditions under which it can be suppressed~\cite{trombas1,kamchatnov2008stabilization,ma2010controlling} in the realm of atomic condensates.
Indeed, the relevant phenomenology has also been long
experimentally observed in the realm of condensates~\cite{Anderson_watching_2001}
and has been also of extensive interest (including experimentally)
in the context of nonlinear optics~\cite{KIVSHAR2000117}.
Yet, it is intriguing to investigate if such an instability can be triggered in a droplet setting, an outcome that is not 
a priori evident given the competing nature of the relevant
nonlinearities.  

Additionally, different types of bound states such as kinks~\cite{kartashov2022spinor} and the so-called bubbles~\cite{katsimiga2023interactions,edmonds2023dark} have been recently reported as 1D states existing at comparatively more negative chemical potentials than droplets. 
These correspond, respectively, to heteroclinic and homoclinic solutions of the underlying phase space. 
In contrast to droplets, bubbles have been
found in such 1D settings to generically constitute unstable structures featuring core expansion and
mutual attraction when placed close to each other~\cite{katsimiga2023interactions}, while kinks 
have been identified as stable configurations.  
It is another aim of the present work to unravel the possible existence of such entities in 2D droplet setups and simultaneously examine their potential 
robustness in the presence of transverse modes.  

Here, we address the aforementioned open questions resorting to a 2D short-range interacting homonuclear bosonic droplet environment described by a corresponding eGPE~\cite{petrov2016ultradilute,luo2021new}. 
More concretely, by leveraging the uniform background along one spatial direction, we initially reduce the 2D eGPE into an effective 1D model featuring the logarithmic nonlinearity. 
The time-independent 1D version of this model is recast into a Newtonian type of equation, reminiscent of the particle picture within an effective potential utilized in repulsively interacting gases~\cite{kevrekidis2004solitary,kevrekidis2015defocusing}. 
This  is used to unveil the existence of a variety of distinct nonlinear states ---in addition to the well-known
2D droplets--- within suitable ranges of (negative) chemical potentials. 
In increasing chemical potential order, the
identified states arise in the form of bubbles, kinks, droplets and DSSs respectively and suitably generalize their 1D counterparts of Ref.~\cite{katsimiga2023interactions}. 
Concrete parametric regions of existence of the relevant waveforms are showcased. 
Furthermore, importantly, we infer the stability of the above stationary states within a Bogoliubov-de-Gennes (BdG) framework of the genuine 2D eGPE model. 
It is found that kinks and droplets are stable structures, in contrast to the generically unstable bubbles and DSSs. 

Specifically, it is demonstrated that a DSS experiences snake instability via which a chain of interacting vortical structures is progressively formed. 
The growth rate of this instability gradually increases as the chemical potential is tuned all the way to positive values. 
However, suppression of the instability occurs for DSSs closer to their lower boundary of existence. 
On the other hand, bubble destabilization manifests through its initial core expansion, and eventual emission of
gray solitons (subject to transverse breakup in their own
right), in addition to being accompanied by the formation of density ripples suggesting the apparent emission of (nearly) radially symmetric dispersive shock-waves. 
Additionally, lack of stationary states consisting of a DSS embedded in a finite droplet is explicated. 
Here, direct dynamics reveals the formation of two counterpropagating droplet fragments separated by an increasing in width DSS. The entire configuration experiences structural deformations that depend on the flat-top or Gaussian-like shape of the original droplet background. 

Utilizing a variational method that was earlier used to predict the snake instability of DSSs in a Bose gas~\cite{Cisneros_reduced_2019}, we derive a generalized analytical (yet approximate) BdG formula relating the 
wavenumber of DSS modulation to its corresponding frequency
(if stable) or growth rate (if unstable). This describes the unstable nature of stripes  in the droplet-bearing environment
considered herein. 
Good agreement of capturing the slope of the relevant instability growth rate of a DSS estimated by the most unstable eigenfrequencies identified in the BdG spectrum with the variational prediction is revealed for small transverse wavenumbers, i.e., in the long wavelength limit. However, the variational outcome fails at larger wavenumbers, in line
also with the discussion of~\cite{Cisneros_reduced_2019}. 

The structure of this work is organized as follows. 
Section~\ref{sec:theory} introduces the considered 2D 
bosonic mixture and the respective  
eGPE model utilized for the description of droplets and study of nonlinear excitations.  
The reduced 1D effective potential allowing to infer the presence of the different 
nonlinear excitations together with their parametric regions of existence are 
discussed in Sec.~\ref{effective_model}. 
In Sec.~\ref{spectrum} we demonstrate the stability properties of the identified 
nonlinear solutions and elaborate on the dynamical manifestation of their  
ensuing instabilities. 
In Sec.~\ref{dispersion_relation} the transverse stability analysis of a DSS embedded in the 
droplet bearing environment is theoretically approximated
by constructing a relevant variational method whose predictions are compared with the ones extracted from the BdG analysis. 
Sec.~\ref{DDS_droplet} elaborates on the absence of a stationary DSS solution in a 2D droplet and showcases its dynamical response.  
We summarize our findings and discuss future 
perspectives in Sec.~\ref{conclusions}.

\section{Two-dimensional droplet setup}\label{sec:theory}

We employ a 2D box trapped homonuclear bosonic mixture characterized by intracomponent repulsion of effective strength $g_{ \uparrow \uparrow}\neq g_{ \downarrow \downarrow}>0$ and intercomponent attraction with coupling $g_{ \uparrow \downarrow}<0$. 
The components of such a mixture can be experimentally emulated, for instance, by the hyperfine states $\ket{ \uparrow} \equiv \ket{F=1, m_F=-1}$ and $\ket{\downarrow} \equiv \ket{F=1, m_F=0}$ of $^{39}$K as in Ref.~\cite{cabrera2018quantum}. 
Consequently, the interaction strengths, depending on the corresponding three-dimensional (3D) scattering lengths, are tunable via the available Fano-Feshbach resonances~\cite{chin2010feshbach} which e.g. for $^{39}$K are located in the vicinity of the magnetic field $B = 59 \,$G~\cite{Errico_Feshbach_2007,Roy_test_2013,Lysebo_Feshbach_2010,Tanzi_Feshbach_2018}. 
It has been demonstrated that in this magnetic field region, the effective 2D intercomponent attraction, $g_{\uparrow \downarrow}$, marginally exceeds the effective intracomponent mean  repulsion $\sqrt{g_{\uparrow \uparrow} g_{\downarrow \downarrow}}$. 
As such, it holds that $\delta g \equiv g_{\uparrow \downarrow}+\sqrt{g_{\uparrow \uparrow} g_{\downarrow \downarrow}} \lesssim 0$ and therefore quantum droplets can form~\cite{luo2021new,petrov2016ultradilute}.

To achieve the 2D geometry the atoms are constrained in the $x$-$y$ plane due to the presence of a tight harmonic trap along the perpendicular $z$-direction~\cite{Petrov_interatomic_2001,Hadzibabic_two_2011}. 
Across the plane a box potential of length $L_x=L_y\equiv L$ exists which is adequately extensive such that it does not impact the finite droplet configurations. 
We remark that quantum droplets have been so far experimentally realized for the $^{39}$K two-component mixture only in three-dimensions~\cite{cabrera2018quantum,semeghini2018self,ferioli2019collisions,cheiney2018bright}, although experimental
considerations do not suggest any fundamental limitation
that would preclude this from occurring in lower
dimensions~\footnote{This stems, in part, from
discussions with Prof. P. Engels who has realized
a $^{39}$K condensate in his lab.}.

In fact, we restrict our investigations to densities and interactions satisfying, $n_{\uparrow} \sqrt{g_{\uparrow \uparrow}} = n_{\downarrow} \sqrt{g_{\downarrow \downarrow}}$~\cite{petrov2016ultradilute,Ma_quantum_2023,petrov2015quantum,semeghini2018self}. 
Here, $n_{\sigma}$ denotes the $\sigma=\uparrow, \downarrow$ component density. 
In this case, it is known that the genuine two-component system reduces to an effective single-component one~\cite{petrov2016ultradilute} which can be  described by a corresponding eGPE~\cite{tengstrand2019rotating,li2018two,Otajonov_variational_2020,examilioti2020ground}.
The aforementioned constraint holds for the ground state and low-lying excitations of the droplet~\cite{petrov2015quantum}, and it was experimentally verified in three-dimensions~\cite{cabrera2018quantum}. 
Under these considerations, the reduced eGPE reads
\begin{equation}
i \hbar \frac{\partial \Psi}{\partial t} = -\frac{\hbar^2}{2m} \nabla^2 \Psi +\frac{\hbar^2}{m} g |\Psi|^2 \Psi \ln \left(  \frac{|\Psi|^2}{n_0 \sqrt{e}} \right).
\label{Eq:eGPE}
\end{equation}
The interaction strength encapsulating in 2D the combined contribution of the mean-field interactions and the LHY term takes the form
\begin{subequations}
\begin{gather}
g = \frac{4 \pi}{\ln \left(  \frac{a_{\uparrow \downarrow}\sqrt{a_{\uparrow \uparrow}a_{\downarrow \downarrow}}}{a^2_{\uparrow \uparrow} \Delta }  \right)   \ln  \left(   \frac{a_{\uparrow \downarrow}\sqrt{a_{\uparrow \uparrow}a_{\downarrow \downarrow}}}{a^2_{\downarrow \downarrow}\Delta }    \right)},
\label{Eq:Interaction_strength} \\
\Delta = \exp\left\{ -\frac{\ln^2 \left( \frac{a_{\downarrow \downarrow}}{a_{\uparrow \uparrow}}  \right)}{2 \ln \left(   \frac{a^2_{\uparrow \downarrow}}{a_{\uparrow \uparrow} a_{\downarrow \downarrow}}  \right)}  \right \}. \label{Eq:delta_parameter}
\end{gather} 
\end{subequations}
In these expressions, $a_{\sigma \sigma'}$ refer to the 2D scattering lengths within ($\sigma=\sigma'$) or between ($\sigma \neq \sigma'$) the components. 
Moreover, 
$n_0 = \frac{e^{-2 \gamma -3/2}}{\pi a_{\uparrow \downarrow} \sqrt{a_{\uparrow \uparrow} a_{\downarrow \downarrow}}} \Delta \sqrt{
\frac{4 \pi}{g}}$ is the droplet equilibrium density in the thermodynamic limit~\cite{petrov2016ultradilute} and  $\gamma \approx 0.577$ is the Euler-Mascheroni constant~\cite{abramowitz_handbook_1972}. 
For completeness, we remark that the relation between the 2D scattering lengths and their three-dimensional counterparts, $a^{(3D)}_{\sigma \sigma'}$, is given by~\cite{Petrov_interatomic_2001,Petrov_Bose_2000}
\begin{equation}
a_{\sigma \sigma'} = 2e^{-\gamma} a_{\perp} \sqrt{\frac{\pi}{0.9}} \exp  \left \{   -\frac{a_{\perp} \sqrt{\pi}}{a^{(3D)}_{\sigma \sigma'} \sqrt{2}}  \right  \}.
\label{Eq:Scattering_length_mapping}
\end{equation}
Notice that a peculiarity of the 2D geometry is that quantum droplets can be also hosted for $\delta a^{(3D)}=a^{(3D)}_{\uparrow \downarrow} + \sqrt{a^{(3D)}_{\uparrow \uparrow} a^{(3D)}_{\downarrow \downarrow}} \gtrsim 0$~\cite{petrov2016ultradilute}. This is in contrast to three-dimensions where, in this region, a gas phase takes place; a result that stems from the dimensional dependence of the LHY contribution~\cite{zin2018quantum}. 

Rescaling the time and length scales in terms of $m/(g n_0 \hbar \sqrt{e})$ and $\sqrt{g n_0 \sqrt{e}}^{-1}$ respectively~\cite{Jalm_dynamical_2019} and normalizing the macroscopic wave function to the total particle number, $\int |\Psi|^2~dx dy =gN=g(N_{\uparrow}+ N_{\downarrow})$ we arrive at the respective dimensionless eGPE 
\begin{equation}
i \frac{\partial \Psi}{\partial t} = -\frac{1}{2} \nabla^2 \Psi +  |\Psi|^2 \Psi \ln\left(|\Psi|^2\right).
\label{Eq:dimensionless_eGPE}
\end{equation}
Henceforth,  all quantities are dimensionless unless stated otherwise.
It is important to recognize that the curious nature of
the logarithmic nonlinearity encompasses in a single term
the competition between MF and LHY-type effects. In particular,
at low densities, the negativity of the logarithm effectively
translates into a focusing (attractive) 
type effect, while at large
densities, the corresponding positivity amounts to a
defocusing (repulsive) nonlinearity.

For the stationary states, to be presented below, we solve the time-independent version of Eq.~(\ref{Eq:dimensionless_eGPE}) relying on a fixed point iterative Newton scheme~\cite{kelley2003solving}. Throughout our investigation homogeneous von Neumann boundary conditions are deployed which are suitable especially for the asymptotics of the bubble and kink states.
On the other hand, for the real-time evolution a fourth-order Runge-Kutta integrator is utilized supported by a second-order finite differences method (cross-checked with a pseudospectral method) to take into account the spatial derivatives. Characteristic  values of the temporal and spatial discretization used are $dt=0.0001$ and $dx=dy=0.06$ respectively.  

It should be noted that typical box sizes $L=300$ in the dimensionless units adopted herein correspond to $564 \, \mu m$.
Also, evolution times of the order of $t \sim 500$ refer to $1.14$ sec in physical units upon considering a transverse oscillator length $a_{z}=0.22 \, \mu m$ ($\omega_{z} = 2\pi \times 5.2 \, \rm{kHz}$).

\section{Effective description $\&$ stationary states}\label{effective_model}

\begin{figure*}
\centering
\includegraphics[width= \textwidth]{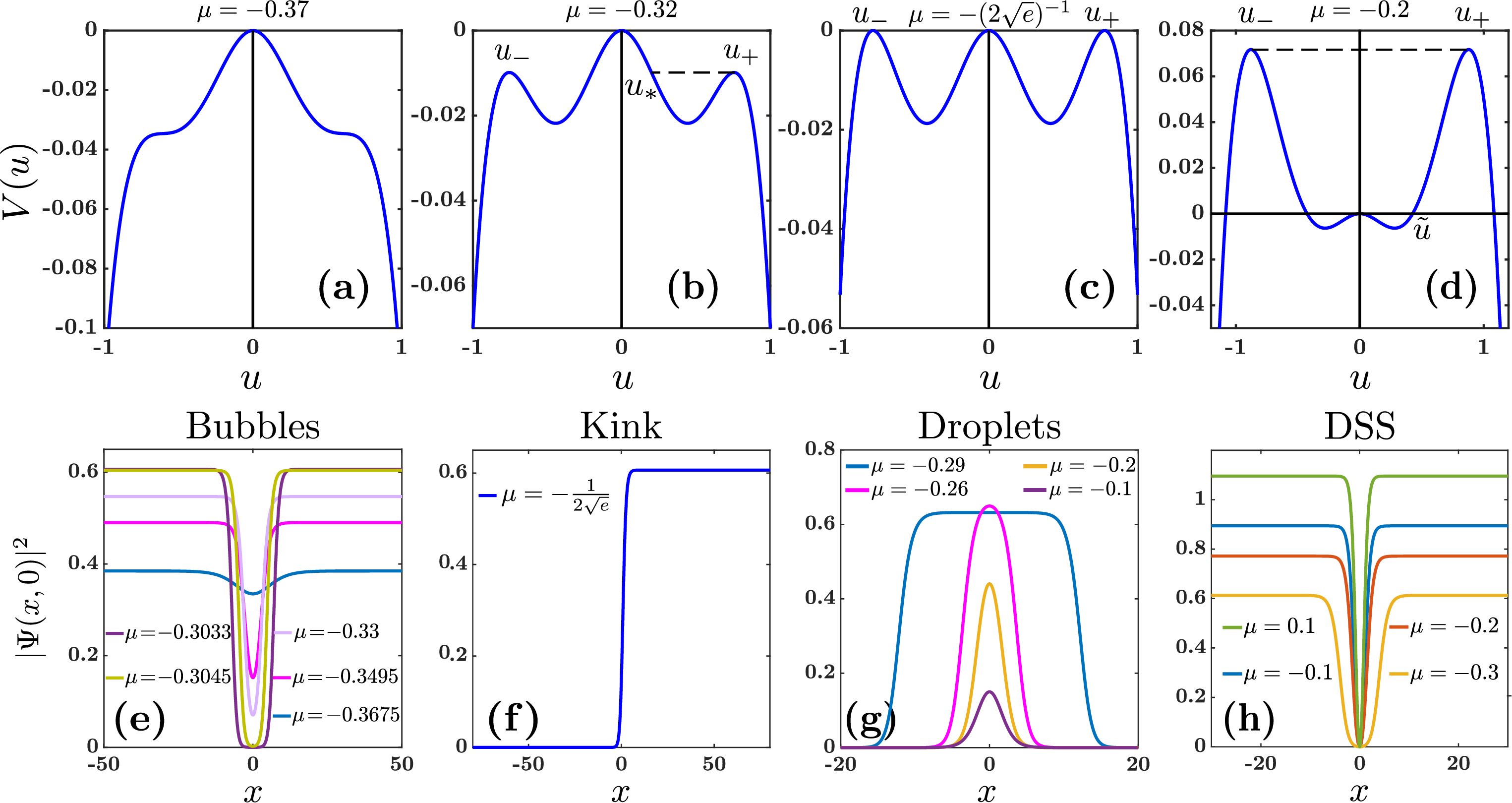}\caption{Quasi-1D profiles 
of the effective potential, $V(u)$, for distinct chemical potential values (see legend). 
(a) For $\mu<\mu^{(1)}_{c}=-0.3679$ no coherent structures are present in the system. (b) Bubbles occur for $\mu^{(1)}_{c}<\mu<\mu^{(2)}_{c}=-1/(2\sqrt{e})$. (c) A kink solution is identified 
for $\mu=\mu^{(2)}_c$. (d) Self-bound quantum droplets and dark solitons coexist in the interval $0>\mu>\mu^{(2)}_{c}$. 
Selected density profiles, $|\Psi(x,0)|^2$, of the different numerically obtained 2D stationary states of the eGPE of Eq.~(\ref{Eq:dimensionless_eGPE}). 
(e) Bubble states progressively acquire wider cores tending to infinitely broad entities towards their lower boundary of existence, $\mu^{(2)}_{c}$.
(f) The kink structure appears only for $\mu=\mu^{(2)}_{c}$.
(g) Droplet solutions are seen to deform from Gaussian-like profiles towards flat-top ones for progressively more negative $\mu$ values.
(h) DSS solutions extend beyond $\mu>0$ becoming gradually wider (and supported on a shallower background) as $\mu$ decreases while being highly localized (on top of a taller background) for positive chemical potential values (see legend). In contrast to bubbles, these states display a $\pi$ phase jump across their core (see also Fig.~\ref{Fig:Instability_dynamics}(b1),(d1)) and their width is significantly smaller compared to the one of bubbles. All parameters are dimensionless. }\label{Fig:Effective_potential}
\end{figure*}

When searching for stationary nonlinear solutions whose density profile extends primarily in 1D (while remaining uniform
along the transverse direction), the reduction of the 2D eGPE [Eq.~\eqref{Eq:dimensionless_eGPE}] to a quasi-1D equation serves as a guide for identifying parameter regimes where such structures may exist. 
To do so, we assume the following ansatz, $\Psi(x,y,t) = u(x) v(y) e^{-i \mu t}$, where $\mu$ is the chemical potential. Also, $u(x)$ stands for the stationary solution along the elongated direction and $\abs{v(y)}^2=\rm{constant}$ is a uniform function in the $y$ direction which can then be absorbed in $u(x)$. 
This solution, without loss of generality, is taken to be real. 
It is important to appreciate
that one-dimensional solutions are
{\it always} trivially also solutions
of the two-dimensional problem and then
the fundamental question is that of their
potential stability or instability under
transverse perturbations which is what we
are exploring herein.
With this ansatz the 2D eGPE reduces to the following quasi-1D equation featuring a logarithmic nonlinearity,
\begin{equation}
\mu u(x) = -\frac{1}{2} \frac{d^2u}{dx^2}  +u^3(x) \ln \big(u^2(x)\big).
\label{Eq:Q1D_reduction}
\end{equation}
The above equation can be rearranged, such that it resembles a classical Newtonian particle of unit mass in an effective
potential $V(u)$ in the form $\frac{d^2u}{dx^2} = -\frac{d V(u)}{du}$. 
The consideration of this effective
potential problem will enable us to identify
systematically the quasi-one-dimensional
equilibrium states of the system some of 
which ---to the best of our knowledge---
have not been previously identified.

Specifically, the aforementioned potential $V(u)$ reads
\begin{equation}
V(u) = \mu u^2 -\frac{u^4}{2} \ln \left(  \frac{u^2}{\sqrt{e}} \right).
\label{Eq:Effective_potential}
\end{equation}
The chemical potential $\mu$ dictates the number and form of the extremal points of $V(u)$ [Fig.~\ref{Fig:Effective_potential}(a)-(d)]. 
In particular, for $\mu<-0.367879=\mu^{(1)}_c$ [see e.g. Fig.~\ref{Fig:Effective_potential}(a)] no minima exist, and thus no stationary solutions are expected to appear. Increasing $\mu$ past this critical value, two symmetric potential wells arise [Fig.~\ref{Fig:Effective_potential}(b)], along with a global maximum at $u=0$ and two symmetrically placed local maxima $u_+$ and $u_-$. 
Here, stationary solutions called bubbles are identified [Fig.~\ref{Fig:Effective_potential}(e)]. 
They are restricted between two positive $u$--values, namely $u_+$ and $u_*$, as indicated by the black dashed line in Fig.~\ref{Fig:Effective_potential}(b). 
These two $u$-values are close to each other as long as $\mu \sim \mu^{(1)}_c$,
and tend further apart while simultaneously $V(u_*)$ and $V(u_+)$ increase as $\mu \to \mu^{(2)}_c=-1/(2\sqrt{e})$ from below.
This deformation of $V(u)$ dictates the region of existence of bubble states, i.e. $\mu^{(1)}_c < \mu < \mu^{(2)}_c$. 

The corresponding density profiles of such bubble configurations obtained as stationary solutions of the time-independent 2D eGPE [Eq.~\eqref{Eq:dimensionless_eGPE}] are depicted in Fig.~\ref{Fig:Effective_potential}(e) for different $\mu$ values,
spanning their entire region of existence. 
For visualization purposes, $\abs{\Psi(x,0)}^2$ is displayed, featuring a density dip in the vicinity of $x=0$, while being on top of a finite background. 
Despite their structural similarity to dark solitons, it
is important to appreciate that such solutions do not exhibit a phase jump and show a prominent width.
These 2D bubble entities, being identified here for the first time, are the siblings of their 1D counterparts discussed in~\cite{katsimiga2023interactions,katsimiga2023solitary}.
This is in line with what has been well-known for
such waveforms in different classes of nonlinear Schr{\"o}dinger models of relevance, e.g.,
to nonlinear optics~\cite{debouard,baras1989}.
As it can be seen from Fig.~\ref{Fig:Effective_potential}(e), they become progressively deeper and wider as $\mu \to \mu^{(2)}_c$, with their depth delineated by the $u_+-u_*$ difference in the effective potential picture.

\begin{figure}
\centering \includegraphics[width= \columnwidth]{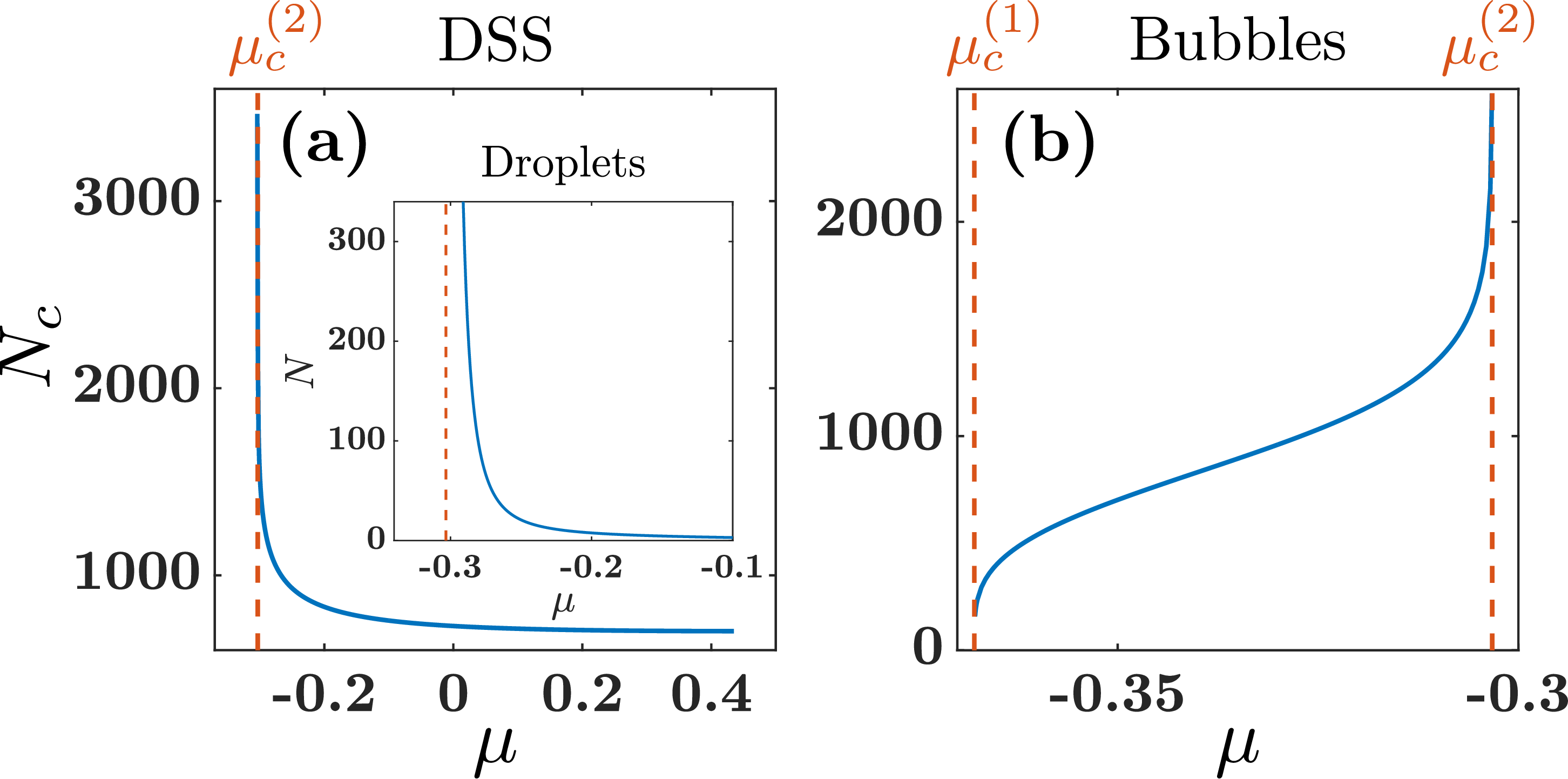}
\caption{Complementary particle number, $N_c$,
of (a) DSS and (b) bubble solutions indicating their regions of existence upon chemical potential variations for $g=1$. Inset in panel (a) presents the particle number, $N$, vs $\mu$ of self-bound droplet solutions. In all cases vertical dashed lines indicate the boundaries of existence of each distinct solution. The lower bound of the droplet and DSS coincide, while the bubble zone starts at $\mu_c^{(2)}$, where droplets cease to  occur, extending to more negative $\mu$, up to $\mu^{(1)}_c$.
Kinks are ``special'' within such a diagram as they 
{\it only} exist at the vertical dashed line point with 
$\mu=\mu_c^{(2)}$. The chemical potential outlined here is unitless. }
\label{Fig:Particle_number}
\end{figure}

For $\mu=\mu^{(2)}_c$, the above-mentioned local maxima $V(u_+)$ and $V(u_-)$, become zero, and thus equal the global one $V(0)$, see Fig.~\ref{Fig:Effective_potential}(c).
Here, the respective classical trajectory connects $u_+ \simeq 0.77$ and $u=0$, implying  that the wave function $u(x)$ is a heteroclinic orbit connecting these two values. 
The relevant density profile satisfying the time-independent version of Eq.~\eqref{Eq:dimensionless_eGPE} refers, accordingly, to a kink configuration, which is in essence a domain wall. It displays an abrupt density jump interlinking a zero background (at $x<0$) to a finite one at $x>0$ as illustrated in Fig.~\ref{Fig:Effective_potential}(f). The values of these backgrounds are in line with the effective potential predictions, $0$ and $u_+$.
We remark that a kink exists also in the relevant 1D eGPE model (and has been found to be stable therein)~\cite{katsimiga2023solitary,katsimiga2023interactions}.

A further increase of $ \mu^{(2)}_c <\mu <0$, leads to pronounced positive valued $V(u_-)$ and $V(u_+)$, while the previous global maximum transitions to a local one [see Fig.~\ref{Fig:Effective_potential}(d)]. 
In this regime, the effective potential can host two kinds of 1D solutions uniformly extending along the transverse direction.
For the energetically lowest lying one $u(x)$ is limited between a finite background $\tilde{u}$ and zero, see Fig.~\ref{Fig:Effective_potential}(d). 
In fact, this solution corresponds to a 1D droplet structure~\cite{petrov2016ultradilute,luo2021new} which is
homogeneous along the transverse direction and obeys Eq. \eqref{Eq:Q1D_reduction} with the logarithmic nonlinearity. 
It gradually deforms upon chemical potential variations from a gaussian ($\mu \to 0$) to a flat-top state ($\mu \to \mu^{(2)}_c$), characterized by exponentially decaying tails at large $\abs{x}$.
The respective $\abs{\Psi(x,0)}^2$ of 
such stationary states extracted from the time-independent 2D eGPE are presented in Fig.~\ref{Fig:Effective_potential}(g).
Notice that at $\mu = \mu^{(2)}_c$, the thermodynamic limit is reached (see also the inset of Fig.~\ref{Fig:Particle_number}(a)), whereas $\tilde{u} = e^{-1/4} \simeq 0.77$~\cite{sturmer2021breathing} corresponds to the equilibrium density predicted by the effective potential.
The interval of existence as well as the shape deformation of droplets can be inferred by monitoring the ensuing particle number, $N= \int dx dy \: \abs{\Psi(x,y)}^2$, with $g=1$ without loss of generality. 
Indeed, close to the thermodynamic limit, i.e. $\mu \to \mu^{(2)}_c$ [vertical dashed line in the inset of Fig.~\ref{Fig:Particle_number}(a)], $N$ increases leading to wider planar droplets~\cite{sturmer2021breathing}, maintaining a (wide region of) nearly constant flat-top density around $n_0$ [Fig.~\ref{Fig:Effective_potential}(g)]. 
It is worth mentioning that the lower bound of the chemical potential in the thermodynamic limit is $\mu^{(2)}_c = -1/(2\sqrt{e})$, which coincides with our numerical predictions.
On the other hand as $\mu \to 0$, solely Gaussian-shaped droplets with comparatively smaller particle number are sustained, eventually ceasing to exist [inset of Fig.~\ref{Fig:Particle_number}(a)].

The energetically higher-lying second kind of solutions within $ \mu^{(2)}_c < \mu <0$ is bounded by $u_+$ and $u_-$, see the black dashed line in Fig.~\ref{Fig:Effective_potential}(d).
More concretely, the wave function has a heteroclinic form
(reminiscent, but not equal to the familiar tanh-shaped black
soliton of the standard cubic 
nonlinear Schr{\"o}dinger model) which
changes sign between the two equal in magnitude but opposite in sign backgrounds.
The stationary excitation associated with this $V(u)$ is the DSS [Fig.~\ref{Fig:Effective_potential}(h) for $\mu <0$]. 
Recall that the latter is a well known nonlinear structure existing in repulsive gases~\cite{frantzeskakis2010dark}, but constitutes a suitably modified entity as per the 2D eGPE model under consideration.
These fully-dipped (down to $0$ in their density) configurations existing on top of a homogeneous background exhibit narrower cores as $\mu \to 0^-$. 
Turning to $\mu>0$, the aforementioned local maximum transitions to a global minimum (not shown), resulting in DSSs that are progressively more localized, as shown in Fig.~\ref{Fig:Effective_potential}(d). 
In line with the effective potential outcome the related background shifts to larger values when compared to $\mu <0$  [Fig.~\ref{Fig:Effective_potential}(h)].

Since both DSS and bubbles are characterized by particle depletion across their cores,
the dependence of their width on $\mu$ variations is captured through the complementary particle number
\begin{equation}
N_c = \int dx dy \:  \left[  \abs{\Psi(x \to \infty,y )}^2 - \abs{\Psi(x,y)}^2  \right ].
\label{Eq:Comp_number}
\end{equation}
Here, $\abs{\Psi(x \to \infty,y)}^2$ refers to the finite DSS or bubble background at $g=1$. 
Focusing on the DSS solutions we observe that for $\mu \to \mu^{(2)}_c$ their cores become wider, and thus $N_c$ diverges, while for increasing $\mu$ towards positive values, their width (background) is narrower (larger), reflected by the saturated $N_c$ [Fig.~\ref{Fig:Particle_number}(a)].
Turning to bubbles, their zone of existence can also be inferred from $N_c$, see vertical dashed lines in Fig.~\ref{Fig:Particle_number}(b).
Concretely, the bubble is shallower (deeper) as $\mu \to \mu^{(1)}_c$ ($\mu \to \mu^{(2)}_c$), implying that
 $N_c \to 0$ ($N_c \to \infty$) [Fig.~\ref{Fig:Effective_potential}(e)].

\section{Spectrum $\&$ dynamics of nonlinear solutions}\label{spectrum}

Having identified the plethora of different possible nonlinear structures that emerge in an effective 1D form within the 2D eGPE model, we subsequently explore their excitation spectra and unstable dynamics.

\begin{figure*}
\centering
\includegraphics[width=\textwidth]{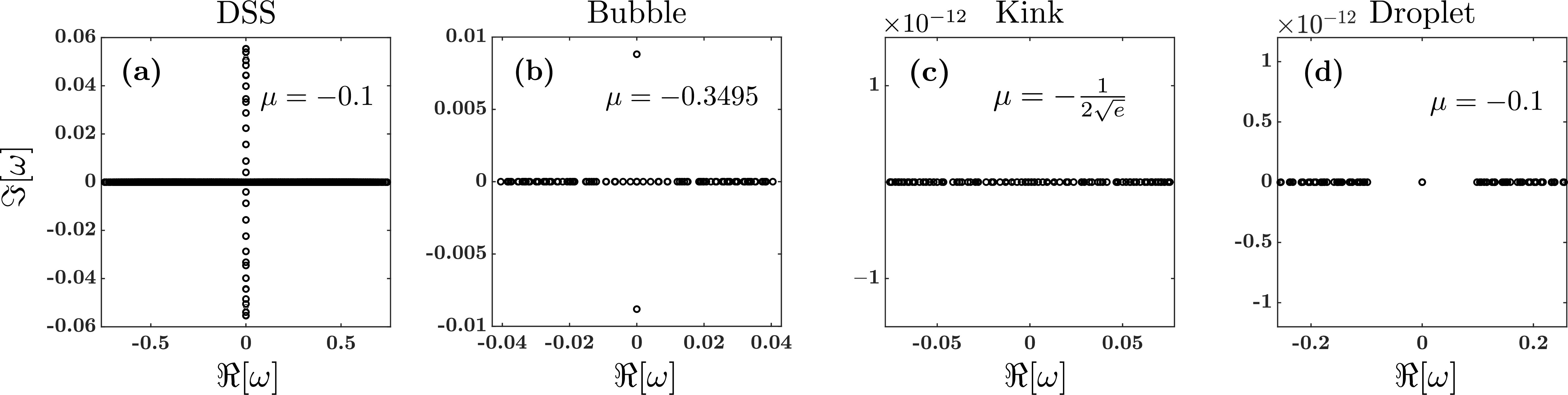}\caption{Selected 2D BdG spectra of the ensuing waveforms demonstrating their stability properties. The existence of a finite imaginary part, $\Im[\omega]\neq 0$, designates the unstable nature of the relevant stationary state. (a) DSSs and (b) bubbles are genuinely unstable configurations in their respective domains of existence, whereas (c) the kink and (d) the droplets are stable solutions throughout their interval of existence. The eigenvalues and chemical potentials are measured in dimensionless units. }
\label{Fig:Profiles_BdG}
\end{figure*}

\subsection{Stability properties of nonlinear states}\label{stability}

The stability of the above-discussed solutions is
addressed by means of a BdG linearization 
analysis~\cite{li2018two}. 
In this context, each stationary state is 
perturbed through the ansatz

\begin{equation}
\Psi(x,y)= \Big(\Psi_0(x,y)+\epsilon a(x,y)e^{-i\omega t}+\epsilon b^{\star}(x,y)e^{i\omega^{\star} t}\Big)e^{-i\mu t},
\label{pert}
\end{equation}
where $\epsilon$ is a formally small perturbation 
parameter. 
Also, $a(x,y)$, $b(x,y)$ denote the relevant eigenvectors and $\omega$ the 
eigenfrequencies. 
Assuming that the wave function is real, and Taylor 
expanding the logarithmic contribution of Eq.~\eqref{Eq:dimensionless_eGPE} while keeping terms 
of order $\mathcal{O}(\epsilon)$, one arrives to 
the following eigenvalue problem

\begin{gather}
\begin{bmatrix}
\hat{L}_{11} & \hat{L}_{12}\\
\hat{L}_{21} & \hat{L}_{22}
\end{bmatrix}
\begin{bmatrix}
a \\
b 
\end{bmatrix}
= \omega
\begin{bmatrix}
a \\
b
\end{bmatrix}.
\label{bdg}
\end{gather}
Here, $\hat{L}_{11}=-\hat{L}_{22}=-\frac{1}{2}\nabla^2-\mu+\Psi^2_0+2|\Psi_0|^2 \rm{ln}(\Psi^2_0)$ and $\hat{L}_{12}=-\hat{L}_{21}=\Psi^2_0+\Psi^2_0 \rm{ln}(\Psi^2_0)$, while
$\nabla^2\equiv \partial^2_x+\partial^2_y$ is the 
2D Laplace operator.

It turns out that DSSs are unstable 
solutions throughout their interval of existence.
This outcome can be inferred by inspecting the 2D BdG spectra presented in Fig.~\ref{Fig:Profiles_BdG}
(a) e.g., for $\mu=-0.1$ where the real, 
$\Re[\omega]$, versus the imaginary, $\Im[\omega]$,
is illustrated.
It is the presence of a finite $\Im [\omega]$ that signals the destabilization of the relevant waveform, as dictated by the first 14 eigenfrequency pairs presented herein.
As we will explicate in detail below, dynamical destabilization occurs due to the exposure of these entities to transverse excitations amplified via the so-called snake instability~\cite{frantzeskakis2010dark,Anderson_watching_2001,KIVSHAR2000117} also within the 2D eGPE model under consideration. 

Similarly to the DSS, 2D bubble configurations are found to be genuinely unstable solutions for all values of $\mu$ spanning their domain of existence.  
This result aligns with the unstable nature of bubbles in the relevant extended 1D model~\cite{katsimiga2020observation}.
A characteristic BdG spectrum is shown in
Fig.~\ref{Fig:Profiles_BdG}(b) 
corresponding to a bubble solution with $\mu=-0.3495$, namely close to the lower bound of existence [Fig.~\ref{Fig:Effective_potential}(e)] of this relatively shallow waveform. 
This bubble's spectrum entails the presence of a single eigenfrequency pair having $(\Re[\omega], \Im[\omega]) =(0, 0.009)$. 
Note however, that as $\mu \to \mu^{(2)}_c$, i.e. for deeper and wider bubbles, the growth rate of the ensuing instability becomes significantly suppressed. 
For instance, it is found to be 
$(\Re[\omega], \Im[\omega]) =(0, 6 \times 10^{-5})$
for $\mu=-0.3033$.
Therefore, these deep bubble waveforms remain robust for long evolution times, which could potentially facilitate their experimental observation.

Strikingly, and also in contrast to the above stability properties, 
precisely for $\mu=\mu^{(2)}_c$, where the effective potential maxima
become degenerate, spectral stability of the kink entity is observed. 
The latter can be readily seen in Fig.~\ref{Fig:Profiles_BdG}(c) where the absence of a finite imaginary eigenfrequency 
($\Im[\omega]=0$) is evidenced.
To the best of our knowledge, this is a prototypical manifestation of spectral stability of 1D topological defects, like the kink solution, in nonlinear Schr{\"o}dinger type models, upon their exposure to transverse perturbations.
Clearly, it is worthwhile to investigate this remarkable finding more
systematically, yet
we defer such a detailed study for future work.  
As a next step, the ensuing BdG analysis of droplets is examined. 
A case example demonstrating the 
stability of these self-bound states~\cite{li2018two} is shown in Fig.~\ref{Fig:Profiles_BdG}(d) e.g. for $\mu=-0.1$.

\begin{figure*}
\centering
\includegraphics[width=\textwidth]{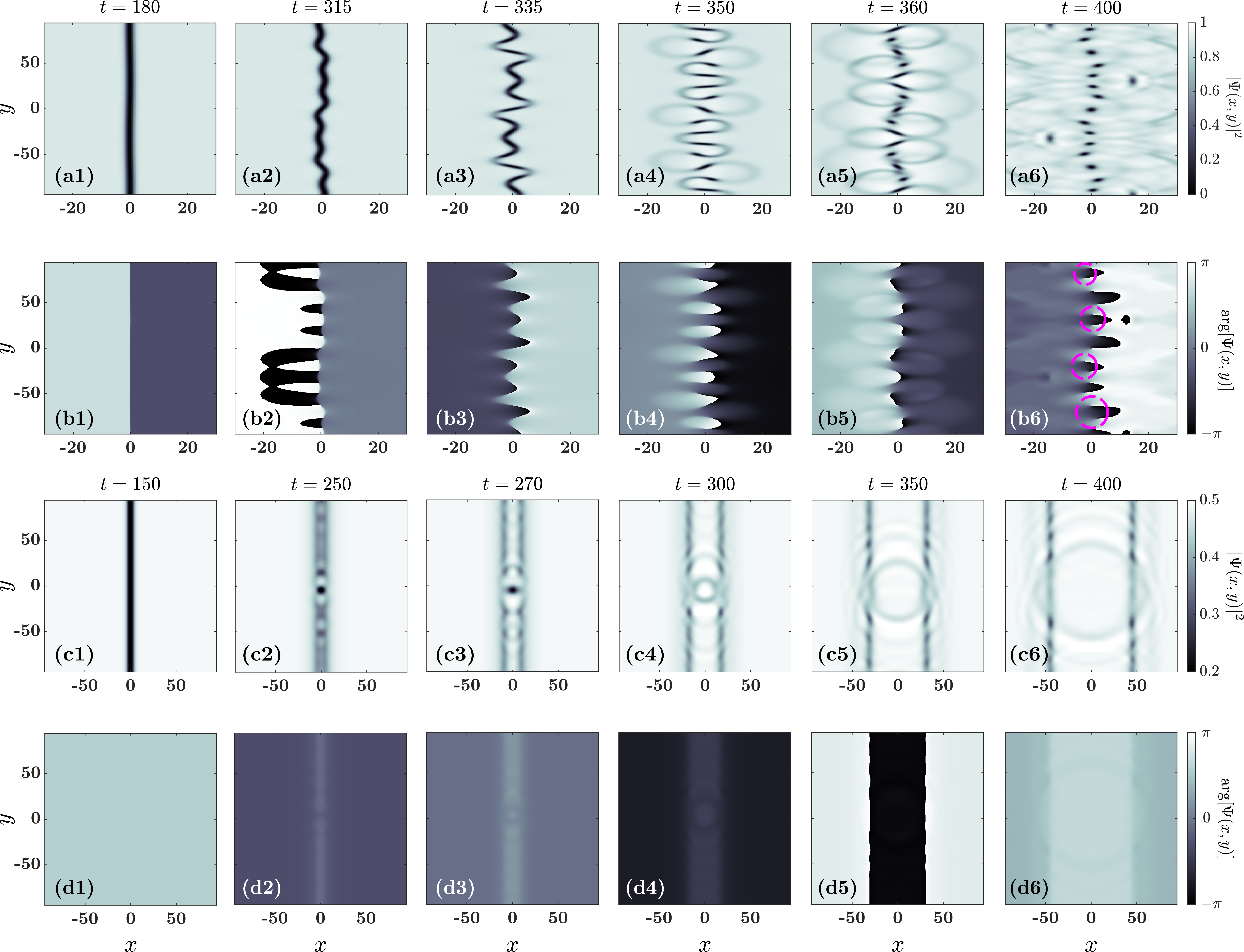}
\caption{Density snapshots, $|\Psi(x,y)|^2$, at different times (see legends) during the dynamical evolution of a $\rm{(a1)-(a6)}$ DSS and $\rm{(c1)-(c6)}$ bubble solution for $\mu=-0.1$ and $\mu=-0.3495$ respectively. 
The corresponding phase profiles of (a1)-(a6) [(c1)-(c6)] are depicted in (b1)-(b6) [(d1)-(d6)]. 
The relevant in each case nonlinear waveform is perturbed by adding to it the eigenvector associated with the most unstable (in the case of the DSS) and single unstable eigenfrequency (for the bubble) mode. 
Snake instability of the DSS manifests itself leading to the break up of the stripe into an array of interacting vortices that progressively covers the entire extent of the initial soliton, see e.g. the developed 2$\pi$ phase jumps in panels (b3)-(b6) denoted with dashed magenta circles. Dynamical destabilization of bubbles is initiated through a core expansion accompanied by the splitting to smaller and shallower bubbles and eventually leading to two counterpropagating stripes. The density ripples appearing in the splitting process are reminiscent of counterpropagating radially symmetric dispersive shock waves. All quantities depicted are given in dimensionsless units. }
\label{Fig:Instability_dynamics}
\end{figure*}

\subsection{Dynamical destabilization}\label{dynamics}

To further shed light on the aforementioned (in)stability analysis findings, we next explore the dynamical evolution of the obtained waveforms. 
Even though our primary focus is to investigate the
destabilization of DSSs and bubbles, we note in passing that we confirmed the robust spatiotemporal evolution of the kink structure and the droplets for times up to $t \sim 2\times 10^3$ in the dimensionless units adopted herein (not shown for brevity). 
This corresponds to $4.5 \, s$
suggesting the relevance of these states and their potential observation in current state-of-the-art experiments.

In contrast to the longevity and coherent evolution of the above entities, the unstable dynamics of DSSs entails their structural deformation. 
Such a progressive shape alteration is depicted in
Fig.~\ref{Fig:Instability_dynamics}$\rm{(a1)-(a6)}$ regarding a DSS obtained for $\mu=-0.1$ and characterized by a $\pi$ phase jump across the $x$-direction [Fig.~\ref{Fig:Instability_dynamics}(b1)].
Particularly, in order to trigger the relevant
destabilization at earlier times, the DSS is perturbed by adding to it the eigenvector associated to the most unstable mode appearing in its BdG spectrum [see Fig.~\ref{Fig:Profiles_BdG}(a)]. 
This mode corresponds to the maximum growth rate of the ensuing instability. 
Around $t\approx 300$, the solitary wave undergoes undulations in the transverse $y$-direction manifesting the initiation of the snake-instability. 
These density undulations progressively expand over the entire extent of the former stripe state [Fig.~\ref{Fig:Instability_dynamics}(a2)-(a4)]. 
This behavior becomes pronounced for $t>300$ [Fig.~\ref{Fig:Instability_dynamics}(a2),(a3)], and it is accompanied by an 
increasing spatial heterogeneity of the phase jump profile [Fig.~\ref{Fig:Instability_dynamics}(b2),(b3)]. Subsequently, transient vortex dipoles start to develop, as captured by the respective $2\pi$ phase jump at their locations [Figs.~\ref{Fig:Instability_dynamics}(b4)]. 
Eventually, vortices of alternating charge appear throughout the original stripe soliton [Fig.~\ref{Fig:Instability_dynamics}(a5)-(a6)]. 
This can be readily seen by the characteristic $2\pi$ phase jumps appearing in the corresponding phase contour plots [Figs.~\ref{Fig:Instability_dynamics}(b5)-(b6)]. 
A similar to the above discussed response occurs upon exciting the DSS with one of lower-lying eigenfrequencies, predicted by the BdG spectrum [Fig.~\ref{Fig:Profiles_BdG}(a)], bearing smaller $\Im [\omega]$. 
The latter naturally implies slower onset of the snake instability (not shown). 
The relevant unstable eigenmodes effectively constitute
a band of unstable perturbations with a corresponding interval
of imaginary eigenfrequencies as will also be demonstrated in section IV 
below.

Far more intriguing is found to be the destabilization of bubbles depicted in Fig.~\ref{Fig:Instability_dynamics}$\rm{(c1)-(c6)}$. 
As in the preceding scenario, also here, the wave function is initially perturbed through the eigenvector associated with the single unstable mode present in the BdG spectrum of Fig.~\ref{Fig:Profiles_BdG}(b).
Evidently, at the early stages of the dynamics bubble core expansion takes place leading to a splitting of the initial waveform into an array of 
gray solitary wave structures 
[Fig.~\ref{Fig:Instability_dynamics}(c2)]. This process is accompanied by the creation of density ripples, in addition to the definitive emergence
of phase jumps associated with the emerging gray solitary structures.
Moreover, the resulting density ripples are reminiscent of counterpropagating (nearly) radially symmetric dispersive shock waves in analogy to their 1D counterparts~\cite{katsimiga2023interactions}, and certainly deserve a separate investigation to understand their nucleation and characteristics.
Each of the newly formed gray solitary structures propagate with their
respective (opposite) speeds, while themselves becoming subject progressively
to a weaker (given their gray nature) transverse instability
[Fig.~\ref{Fig:Instability_dynamics}(c3)-(c6)].
Inspecting the phase profiles during dynamics, reveals that the original uniform phase [Fig.~\ref{Fig:Instability_dynamics}(d1)] experiences,
in addition to the gray soliton phase jumps, slight local 
undulations along the solitonic gray stripe in the course of the evolution, see Figs. \ref{Fig:Instability_dynamics}(d2)-(d6).

\begin{figure*}
\centering
\includegraphics[width= \textwidth]{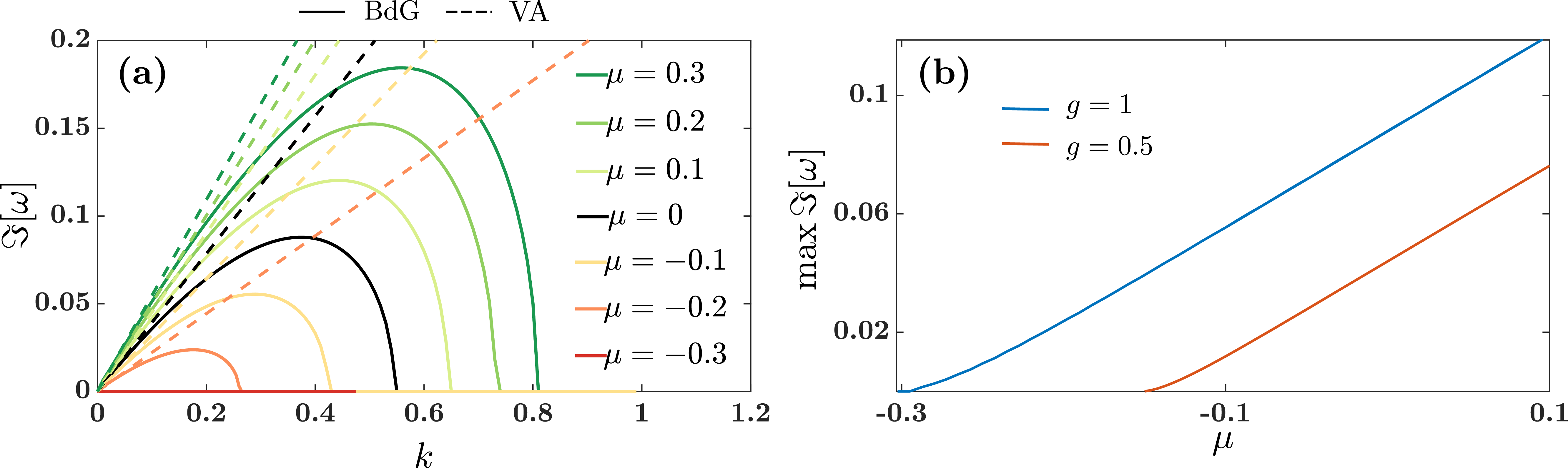}\caption{(a) Stability spectra of the DSS solution in a homogeneous background for different chemical potentials spanning its entire domain of existence (see legend). The imaginary eigenfrequency, 
$\Im[\omega]$, i.e., the growth rate of the corresponding
instability obtained from the BdG as a function of the wavenumber, $k$, is illustrated. In all cases dashed lines correspond to the analytical prediction  obtained through the variational approach [see Eq \eqref{Eq:Dispersion_relation}]. The latter estimates a linear growth of the instability rates as $k$ increases that matches nicely with the slope obtained by our 2D BdG numerics for small $k$ values, i.e., at the long-wavelength
limit. However, for increasing $k$ the BdG shows a different behavior compared to the variational result  for the instability branches which acquire a maximum value for each $\mu$, in a way
similar to what is known for the cubic defocusing NLS
model.
(b) Maximum growth rate, $\rm{max} \Im[\omega]$, upon $\mu$ variations showcasing its nearly linear increase for different coupling constants (see legend). All variables are provided in dimensionless units. }
\label{Fig:Dispersion}
\end{figure*}

\section{Stability Analysis of a DSS in the 2D eGPE}\label{dispersion_relation}

Next, we aim to investigate further the destabilization of the DSS in this 
environment featuring competing nonlinearities.
To this end, we generalize a variational approach (VA) that was employed to analytically probe the excitation spectra of multiple DSSs embedded in 2D repulsive condensates~\cite{Cisneros_reduced_2019}.
In this context, it was argued that the VA predictions display good qualitative agreement with the numerical BdG outcome.

More concretely, motivated by the observed dislocation of the DSS center associated with the snake instability [Figs. \ref{Fig:Instability_dynamics}(a1)-(a6)], the following variational ansatz is employed~\cite{Cisneros_reduced_2019,Kivshar_lagrangian_1995},
\begin{gather}
\Psi^{(\rm{VA})}(x,y,t)= B(y,t) \tanh \left[ D(y,t)(x-X(y,t)) \right] \nonumber \\
+ iA(y,t).
\label{Eq:Variational_ansatz}
\end{gather}
Evidently, three independent variational variables are introduced, namely the position of the stripe $X(y,t)$, the soliton inverse width $D(y,t)$, and the soliton 
depth parameter $B(y,t)$.
All of them depend explicitly on time and the $y$ coordinate, since the DSS destabilization at the initial dynamical stages primarily occurs along this direction [see e.g. Fig.~\ref{Fig:Instability_dynamics}(a2)] due to the presence of transverse modes. 
The velocity-associated quantity $A(y,t)$ is related to the original DSS background $u_0=B(y,0)$, through $A^2(y,t) + B^2(y,t) = u_0^2$. 
For the well-established case of the cubic
repulsive nonlinearity, it is known that in the case of a dark soliton its inverse width is equal to its depth parameter $B$~\cite{frantzeskakis2010dark}. 
Moreover, utilizing the above VA it was found that $D(y,t)=B(y,t)$ adequately holds also for relatively small perturbations on both the DSS width and center across the transverse direction~\cite{Cisneros_reduced_2019}.

However, for the eGPE considered here, our simulations indicate that such a condition is approximately met only at large positive chemical potentials. 
In sharp contrast, at negative $\mu$, the inverse width becomes smaller than the background [Fig.~\ref{Fig:Effective_potential}(h)]. 
Recall that in both cases, $\mu=u_0^2 \ln(u_0^2)$ for a uniform background $u_0$ , see also Eq.~\eqref{Eq:dimensionless_eGPE}.
In fact, according to our eGPE simulations for varying $\mu$ it turns out that the inverse width with respect to $u_0$ in the interval $-0.3 < \mu \lesssim 0$ can be best approximated by $a\ln(u_0-b) +cu_0 +d$, where $a$, $b$, $c$, $d$ are fitting parameters\footnote{Based on the effective quasi-1D equation [Eq. \eqref{Eq:Q1D_reduction}] and assuming $u(x)=u_0 \tanh(Dx)$ and $\mu=u_0^2 \ln(u_0^2)$, it is possible to predict the inverse width, $D=\frac{u_0}{12} \sqrt{9\pi^2-80+12\ln(u_0^2)}$. 
The latter is a good approximation at $\mu>0$, in contrast to the fitted function describing the entire $\mu$ region of DSS existence.}.  
In part this fitting is inspired by the linear behavior for large $u_0$ as anticipated in the limit of large positive chemical potentials~\cite{kevrekidis2015defocusing}, where the defocusing NLS-like behavior
dominates the dynamics. 
These aspects motivate the extension of the variational ansatz introduced in Ref.~\cite{Cisneros_reduced_2019}, where the adapted inverse width reads $D(y,t)=f(B(y,t))=a\ln(B(y,t)-b) +cB(y,t) +d$, holding for the stationary states and small perturbations as well. While this approximation, slaving the
dynamics of the inverse width to those of
the soliton depth, is a priori valid for the stationary
case, we will also justify it a posteriori in what
follows for the dynamically evolving case.

We start from the 2D Lagrangian pertaining to the DSS~\cite{Kivshar_lagrangian_1995,kivshar_dark_1998}
\begin{gather}
L = \int dx dy \: \Bigg \{ -\frac{i}{2} \left( \Psi^* \frac{\partial \Psi}{\partial t} -\Psi \frac{\partial \Psi^*}{\partial t} \right) \left( 1-\frac{u_0^2}{\abs{\Psi}^2}  \right) \nonumber \\
+\frac{1}{2} \abs{\nabla \Psi}^2 +\frac{1}{2} \abs{\Psi}^4 \ln \left( \frac{\abs{\Psi}^2}{\sqrt{e}} \right) -\frac{u_0^4}{2} \ln \left( \frac{u_0^2}{\sqrt{e}}\right)  \nonumber  \\
+u_0^2\ln(u^2_0) \left( u_0^2 -\abs{\Psi}^2  \right)  \Bigg\}.
\label{Eq:Lagrangian}
\end{gather} 
The term $-u_0^2/\abs{\Psi}^2$ ensures the momentum renormalization of the DSS within the Lagrangian formulation known for nonlinear Schr\"odinger type models, without affecting  the equation of motion~\cite{Kivshar_lagrangian_1995}.
In a similar vein, the last two terms serve as renormalization constants, removing divergences that occur due to the finite background $u_0$ extending in the entire $x$-$y$ plane. 

Subsequently, we plug the variational ansatz of Eq.~\eqref{Eq:Variational_ansatz} into the above Lagrangian, and integrate out the $x$-direction, where the soliton lies. As such, it is viable to obtain the reduced $L_y$ Lagrangian with two independent variables, the time $t$ and the $y$ coordinate.
The background on which the localized DSS is mounted remains almost constant during the initial stages of the dynamics [Fig.~\ref{Fig:Instability_dynamics}(a2)-(a3)].
On the contrary, the inverse width and position of the soliton are modified,
allowing us to approximate the $y$ derivative of the wave function as (in a way reminiscent of the corresponding calculation of~\cite{Cisneros_reduced_2019}) 
\begin{gather}
\Psi^{(\rm{VA})}_y \simeq B \left[  f'(B)B_y (x-X) -f(B) X_y  \right ]  \nonumber \\
\times \sech^2 \left[ f(B)(x-X)  \right].
\label{Eq:Derivative_approximation}
\end{gather}
Notice that here we are accounting for the contribution 
of the $y$-variation that remains square-integrable. 
In addition to the above motivation, we will seek to
justify this approximation a posteriori also via its comparison
to the numerical results discussed later in this section.
For simplicity, all variable arguments have been dropped.
Moreover, to derive a closed expression for the integral of the two last terms in Eq.~\eqref{Eq:Lagrangian}, the following trigonometric identity and variable substitutions have been employed,
i.e., $\sech^2(x)+\tanh^2(x)=1$ and $B/u_0=\cos(\phi)$ respectively\footnote{Note that when performing the integrals, one needs to use the formula of the dilogarithms~\cite{maximon_dilogarithm_2003}, $L_2(e^{2i\phi}) +L_2(e^{-2i\phi})=-\pi^2/6+(2\phi-\pi)^2/2$, holding for $\phi \in (0,\pi)$.}.

Upon these considerations, the integrated Lagrangian 
density $L_y$, such that $L=\int dy L_y$  takes the form
(where we have integrated with respect to $x$)
\begin{flalign}
&L_y = \frac{30}{9 f(B)} B^2 u_0^2 -\frac{10}{9} \frac{B^4}{f(B)}    &\nonumber \\
&+\frac{6}{9} \frac{B}{f(B)} u_0^3 \sqrt{1-\frac{B^2}{u_0^2}} \left( -5 +2\frac{B^2}{u_0^2}\right) \arcsin \left( \frac{B}{u_0}  \right)  &\nonumber \\
&-\arcsin^2 \left( \frac{B}{u_0}  \right) \frac{u_0^4}{f(B)} 
+\frac{B}{f(B)} \ln  \left( \frac{u_0^2}{\sqrt{e}} \right) \left[ -2Bu_0^2 +\frac{2}{3} B^3 \right]&   \nonumber \\
&+\frac{2}{3} B^2 f(B) 
+ \frac{2}{3} B^2 f(B) X^2_y 
+ \frac{B^2 B_y^2 [f'(B)]^2}{18 f^3(B)} (\pi^2-6)&   \nonumber \\
&+2X_t \left[ B \sqrt{u_0^2-B^2} -u_0^2 \arctan \left[ \frac{B}{\sqrt{u_0^2-B^2}} \right] \right]&  \nonumber \\
 &+2u_0^2\ln(u_0^2) \frac{B^2}{f(B)} .&
\label{Eq:Explicit_Lagrangian}
\end{flalign}
The equations of motion for the $X$ and $B$ variational parameters are then derived from the Euler-Lagrange equations as a function of the two independent variables $(t,y)$.
To capture the initial stages of the snake instability, we assume small perturbations of the $A,X$ variational 
theory dependent variables around their stationary values, i.e. $A(y,t)=A_0(y,t) +\lambda(y,t)$ and $X(y,t)=X_0(y,t) +\xi(y,t)$, where $\lambda,\xi \ll 1$.
Since the stationary solution is a DSS it holds (without
loss of generality in connection to DSS translations)
that $X_0=A_0=0$.
Consequently, the Euler-Lagrange equations are linearized to first order in $\lambda$ and $\xi$, resulting in the equations of motion for the perturbations 
\begin{subequations}
\begin{flalign}
&\lambda_t = -\frac{u_0}{3} \left[ a\ln(u_0-b)+cu_0+d \right]  \xi_{yy}, & \label{Eq:Linearized_velocity} \\
&\xi_t = \frac{\lambda}{72 u_0^2 \left(a \ln \left(u_0-b\right)+c u_0+d\right)^2}& \nonumber \\
&\times \Bigg\{24 u_0 \left(a \ln \left(u_0-b\right)+c u_0+d\right)^3 &\nonumber \\
&+12 u_0^2 \left(\frac{a}{u_0-b}+c\right) \left(a \ln \left(u_0-b\right)
+c u_0+d\right)^2& \nonumber \\ 
&+16 u_0 \left( 3u_0^2 \ln \left(u_0^2\right)+ u_0^2\right) \Big[a \ln \left(u_0-b\right) +c u_0+d\Big]& \nonumber \\
&+2 \left(\frac{a}{u_0-b}+c\right) \Bigg(-26 u_0^4 -6 u_0^4 \ln \left(u_0^2\right) 
+9 u_0^4 \frac{\pi^2}{4}\Bigg) \Bigg\}.&
\label{Eq:Linearized_position}
\end{flalign}
\end{subequations}
For the linearization, the explicit fitting function $f(B)$ has been employed.

Differentiating Eq.~\eqref{Eq:Linearized_position} with respect to time, and substituting $\lambda_t$ from Eq.~\eqref{Eq:Linearized_velocity}, one arrives at a wave equation describing the propagation of perturbations of the DSS position. 
To emulate the perturbation of the DSS position, as identified in Fig.~\ref{Fig:Instability_dynamics}(a2), we use a plane-wave ansatz $\xi(y,t) = \xi_0 e^{i(ky - \omega t)}$, with transverse wavenumber $k$ and eigenfrequency $\omega$. 
In this way, an expression reflecting the 
transverse stability of the DSS is established determining the stability of the small perturbations 
\begin{flalign}
&\omega^2 = -\frac{   k^2  }{216 u_0 \left(a \ln \left(u_0-b\right)+c u_0+d\right)} & \nonumber \\
&\times \Bigg\{24 u_0 \left(a \ln \left(u_0-b\right)+c u_0+d\right)^3 &\nonumber \\
&+12 u_0^2 \left(\frac{a}{u_0-b}+c\right) \left(a \ln \left(u_0-b\right)
+c u_0+d\right)^2& \nonumber \\
&+16 u_0 \left(3 u_0^2 \ln \left(u_0^2\right)+ u_0^2\right) \Big[a \ln \left(u_0-b\right)
+c u_0+d\Big]& \nonumber \\
&+2 \left(\frac{a}{u_0-b}+c\right) \Bigg(-26 u_0^4 -6 u_0^4 \ln \left(u_0^2\right) 
+9 u_0^4 \frac{\pi^2}{4}\Bigg)  \Bigg\}.&
\label{Eq:Dispersion_relation}
\end{flalign}
As it will be evidenced below such a relation admits imaginary eigenfrequencies $\omega$, manifesting the unstable nature of the DSS.

A direct comparison of the aforementioned VA predictions with the relevant outcome obtained upon numerically solving the eigenvalue problem of Eq.~(\ref{bdg}) is provided in Figure~\ref{Fig:Dispersion}(a).
Specifically, Fig.~\ref{Fig:Dispersion}(a) depicts the spectra of DSS solutions for distinct chemical potentials, $\mu$, covering the region of existence of this configuration that ranges from negative ($\mu>\mu^{(2)}_c$) all the way to positive $\mu$ values. 
The imaginary eigenfrequency, $\Im[\omega]$, as a function of the permitted transverse wavenumber, $k_y\equiv k$, is illustrated, demonstrating the growth rates of instability along with the most unstable wavenumbers, $k_{\rm{max}}$, and the relevant stability cut-offs $k_c$. 
The former wavenumber, $k_{\rm{max}}$, refers to the mode that maximizes the imaginary part of the eigenfrequency $\omega$, i.e. $\rm{max}\Im[\omega]$, whereas the latter  critical wavenumber 
is associated with
$\Im[\omega(k_c)]=0$ and designates the threshold of the unstable region.
Recall that due to the effectively 1D nature of the DSS along with Eq.~(\ref{Eq:dimensionless_eGPE}), 
the transverse perturbation can be decomposed into 
Fourier modes, i.e. $a(x,y) = \int dk_y~ e^{-ik_y y} \tilde{a}(x,k_y)$. Their effect on the stability is assessed
here by solving the eigenvalue problem described by  Eq.~\eqref{bdg} for every transverse wavenumber $k_y$ permitted
by the von Neumann boundary conditions
imposed along the transverse direction
boundary.

It turns out that the parametric window of DSS destabilization becomes wider for progressively more positive chemical potentials being significantly  suppressed as $\mu^{(2)}_c$ is approached [Fig.~\ref{Fig:Dispersion}(a)]. 
Notice that the relevant stability equation of  Eq.~\eqref{Eq:Dispersion_relation} matches nicely the corresponding slope of our stability analysis outcome for small $k$ values [dashed lines in Fig.~\ref{Fig:Dispersion}(a)]. 
Namely, upon suitable fitting, it accurately captures the long wavelength limit
where we expect such a variational theory to be accurate. 
Importantly, it adequately captures the relevant slope for chemical potentials closer to the lower boundary of existence of this waveform generalizing in this way earlier findings appearing in cubic NLS models~\cite{Cisneros_reduced_2019} now within the eGPE framework~\footnote{Of course, it does not elude us
that the work of~\cite{Cisneros_reduced_2019} 
has sought to provide a systematic
description of the instability spectrum throughout the
range of unstable wavenumbers. While we do not address this
aspect within the eGPE herein, it is an interesting
avenue for further exploration}.

\begin{figure*}[t!]
\centering
\includegraphics[width=\textwidth]{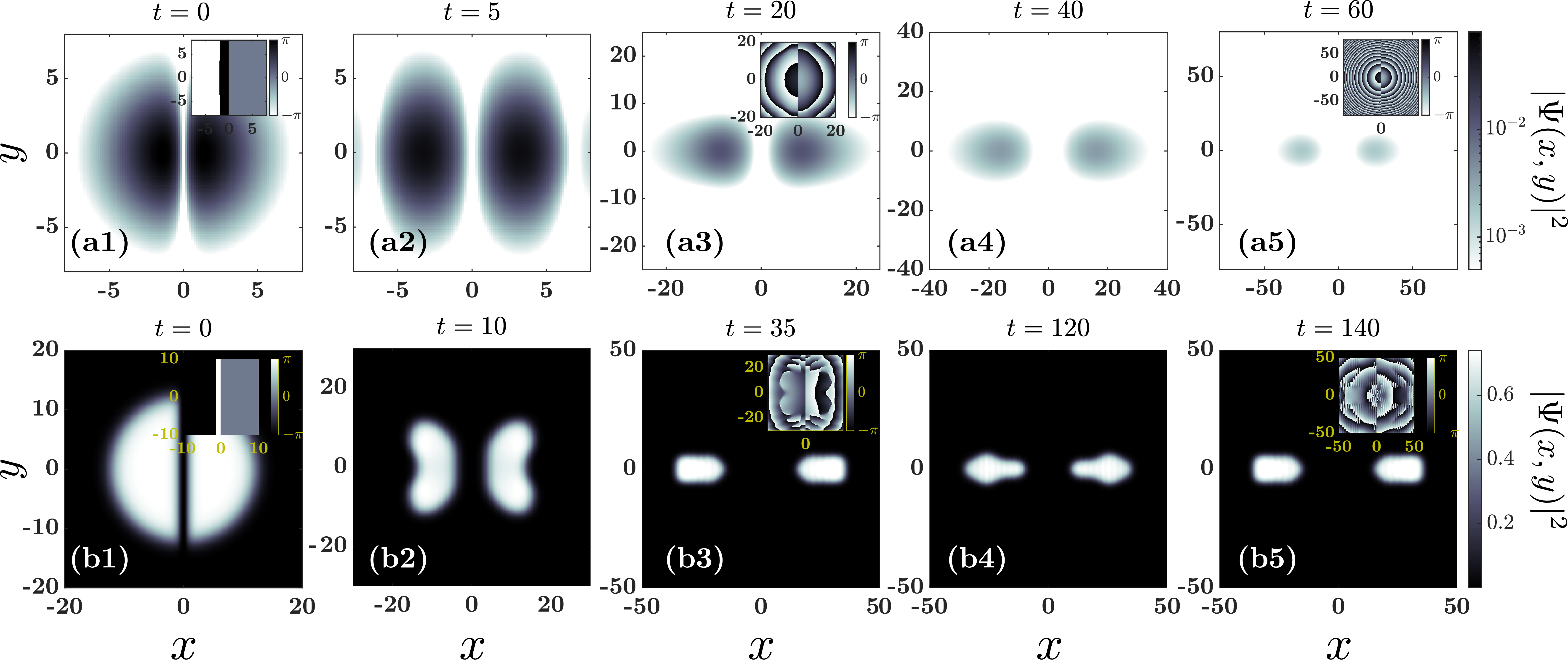}
\caption{Time-evolution of the density, $|\Psi(x,y)|^2$, of a composite structure consisting of a DSS immersed in a finite droplet background with $\rm{(a1)-(a5)}$ $\mu=-0.1$ and $\rm{(b1)-(b5)}$ $\mu=-0.29$. 
Insets depict the respective phase profiles. 
In all cases the unstable dynamics of the DSS in the finite Gaussian ($\mu=-0.1$) or flat-top ($\mu=-0.29$) droplet is observed accompanied by its break up into two-counterpropagating droplet fragments. 
This response suggests the absence of a relevant stationary state. Time and spatial coordinates are depicted in dimensionless units.}
\label{Fig:DSS_droplets}
\end{figure*}

Next, the trend of the different maximal growth rates is monitored for completeness in Fig.~\ref{Fig:Dispersion}(b) upon considering 
$\mu$ variations. In the case of $g=1$ considered throughout the text, a linear growth of the instability of DSS is observed, as
$\mu$ is increased.
Apparently, instability is suppressed at $\mu^{(2)}_c$, which coincides with the thermodynamic limit value of the chemical potential, while it is linearly enhanced towards positive values of $\mu$. This is in line with the identified stability
at that critical value of the kink-like structure
replacing the DSS in the limit of $\mu \rightarrow \mu^{(2)}_c$.
Furthermore, this evinces that snake instability manifests at smaller timescales for progressively more repulsive interaction
dominated environments.
To unveil the general dependence of the growth rate on the interaction strength we employ a different dimensionalization \footnote{In this case the time and length units are rescaled in terms of $m/(\hbar n_0 \sqrt{e})$ and $\sqrt{n_0 \sqrt{e}}^{-1}$.} of Eq.~\eqref{Eq:dimensionless_eGPE}.
In this sense, we calculate the growth rate exemplarily for $g=0.5$, shown in Fig.~\ref{Fig:Dispersion}(b). As it can be seen, the overall 
phenomenology is persistent for different couplings, but being shifted to more positive chemical potentials. This shift reflects the relevant window of existence of the DSS solution.
It can be also inferred that the growth rate is smaller for reduced interactions and fixed $\mu$.

\section{Dynamical response of a DSS in a 2D droplet}\label{DDS_droplet}

Finally, we have also investigated the possibility of the potential existence of a DSS embedded in a finite Gaussian or flat-top droplet instead of a homogeneous background.
However, despite our efforts, it was not possible to numerically obtain, utilizing a Newton iterative scheme,  relevant stationary states even though, according to the quasi-1D effective potential picture discussed  in Sec.~\ref{effective_model}, both entities coexist in the same $\mu$ interval.
The lack of such stationary states is attributed 
to the energetics of the effective potential of the system, $V(u)$, depicted in Fig.~\ref{Fig:Effective_potential}(d).
Namely, the DSS corresponds to a higher energy
than the droplet
and hence cannot be ``harbored'' inside the latter.
We remark that our exploration involved different chemical potentials of the droplet background and widths of the DSS. For simplicity, below, we discuss only the case of a soliton inverse width $D=3$.

Nevertheless, we attempted to tackle these types of states also dynamically and two case examples of our exhaustive studies showcasing the ensuing evolution are provided in Fig.~\ref{Fig:DSS_droplets}. 
In both cases a DSS is embedded in either a Gaussian-shaped droplet for $\mu=-0.1$ [Fig.~\ref{Fig:DSS_droplets}(a1)-(a5)] or a flat-top background for $\mu=-0.29$ [Fig.~\ref{Fig:DSS_droplets}(b1)-(b5)].
The initial state then reads $\Psi(x,y,t=0) = \Psi_0(x,y) \tanh(Dx)$, where $\Psi_0(x,y)$ is the droplet ground state, while the hyperbolic tangent represents the embedded DSS with inverse width $D$.
In the presence of a Gaussian droplet it can be seen that, from the early stages of the dynamical evolution, the DSS begins to expand its core instead of destabilizing through a snake-instability. This expansion is naturally accompanied by the progressive separation of the initial droplets into two oppositely moving fragments, maintaining their individual droplet character.
Importantly, the DSS character is retained in the course of the evolution as can be inferred from the respective phases exhibiting a $\pi$ phase jump in the location of the original stripe [see insets of Fig.~\ref{Fig:DSS_droplets}(a1),(a3),(a5)].
Naturally, this is mandated by the topological protection
of the relevant phase jump profile.

An alternative example of dynamical response takes place in Fig.~\ref{Fig:DSS_droplets}(b1)-(b5).
Indeed the original configuration [Fig.~\ref{Fig:DSS_droplets}(b1)] bearing the characteristic signature of a DSS, see the corresponding phase provided in the inset of panel (b1), substantially deforms along both spatial directions already at early evolution times [Fig.~\ref{Fig:DSS_droplets}(b2)].
Particularly, the two droplet fragments transversely bend while moving further apart due to the DSS core expansion.
The emergent butterfly-like pattern progressively transforms into two highly elongated density lumps along the $x$-direction [Fig.~\ref{Fig:DSS_droplets}(b3)], which are pushed even further apart in the course of the evolution [Figs. \ref{Fig:DSS_droplets}(b4)-(b5)].
This tendency of the DSSs to acquire wider cores during evolution is reminiscent of the stationary DSS solutions for more negative chemical potentials, when embedded in a homogeneous background [Fig.~\ref{Fig:Effective_potential}(h)].
Interestingly, the above patterns reappear at later times (not shown), in their original form and in a mirrored version along the $x=0$ axis.

\section{Summary \& Perspectives}\label{conclusions} 

In the present work we explored the existence, stability and dynamics of nonlinear waves that can be supported within a 2D extended GPE framework encompassing the competition of 
attractive and repulsive interactions at different densities. 
The relevant model featured a logarithmic nonlinear coupling encapsulating the contribution of both the mean-field interactions and first-order LHY quantum correction.
To understand the origin of the different 
transversely homogeneous nonlinear structures, a quasi-1D effective potential picture was analytically extracted, capturing also their concrete parametric regions of existence.
Specifically, by considering chemical potential variations it was possible to infer the presence of DSSs, bubbles, kinks and droplets. 
These configurations and their intervals of existence were also 
confirmed by solving numerically the 2D eGPE.

The stability properties of the aforementioned configurations were examined utilizing a generalized Bogoliubov-de-Gennes analysis. The latter reveals that DSSs and bubbles are generically unstable solutions throughout their domain of existence. 
Remarkably, kink states that exist only for a specific value of the chemical potential are found to be stable solutions against the presence of transverse modes. 
This important finding certainly merits further investigation since it constitutes, to the best of our knowledge, 
a prototypical dynamically stable one-dimensional topological state that persists upon its exposure to transverse excitations.
Additionally and in line with earlier findings~\cite{li2018two}, the spectral stability of 2D droplets is verified.

Furthermore, in order to shed light on the destabilization of the DSS that is found to be enhanced for positive chemical potentials as compared to negative ones, we developed a variational approach.  
Analytically approximating the stability features of these structures allowed for a direct comparison with the relevant numerically obtained 2D spectra.
Such a comparison revealed an adequate agreement between the variational and the numerical BdG predictions in terms of estimating the actual slopes of the underlying instability growth rates at small transverse wavenumbers, i.e., at the long
wavelength limit.  
However for increasing wavenumber the variational approach fails to capture the dispersion relation, an outcome that is corroborated by the BdG spectrum. The latter topic 
remains an interesting open avenue for future investigation.

Moreover, in all cases the dynamical evolution of the ensuing waveforms served as a confirmation of the identified stability properties, revealing among others, the emergence of snake-instability for a DSS.
This instability leads to the production of alternating
charge multi-vortex patterns within the system. 
Interestingly, a rather intriguing destabilization of bubbles occurs. 
In this context the underlying instability entails the structural deformation of bubbles through an initial core expansion 
and associated production of a pair of gray solitonic structures
accompanied by density wave ripples, 
which suggest the spontaneous 
generation of radially symmetric shock-waves.
The gray solitary waves are themselves subject to 
transverse breakup and the radial or nearly-radial
shock wave patterns are also a topic worthy of
further consideration in future studies. 

Furthermore, we investigated the existence of stationary droplet-DSS states with the droplet background being either Gaussian or flat-top. 
Here, due to the  energy and amplitude difference of the two entities,  identification of such composite configurations did not come to fruition. 
Nevertheless, 
the imprinting of a phase jump within a droplet enabled
the observation of such phase pattern-segregated droplets.
The dynamics of the resulting structures unveiled the break-up of the original droplet into two fragments being separated by a continuously expanding region preserving the imposed phase jump. 
The ensuing structural density deformation was found to crucially depend on the droplets' Gaussian or flat-top character. 

There is a multitude of intriguing pathways, based on our findings, to be pursued in the future. 
A straightforward one would be to extract the stability conditions of moving soliton configurations in 2D. This can be achieved by extending the above-discussed variational method and comparing with the predictions of the  BdG analysis as it was demonstrated in the repulsive gas phase~\cite{Cisneros_reduced_2019}. 
An additional technical (yet 
nontrivial) problem is that 
of developing a variational formulation that
accounts for the transverse direction without
any additional assumptions (such as herein
or in~\cite{Cisneros_reduced_2019} the
assumption of practically constant background). 
Moreover, it would be worthwhile to develop a particle picture as it was recently done in the relevant 1D setup~\cite{katsimiga2023interactions}, in order to study the interactions of at least two dark soliton stripes. Another central direction concerns the further investigation of the remarkable stability of the kink structure in similar higher-dimensional (2D, but also 3D) models from both 
atomic and optical physics settings featuring examples
of competing nonlinearities. 
Additionally, the study of the  stability properties of more complex nonlinear structures, such as dark-bright and vortex-bright solitons, that may appear in heteronuclear genuine two-component droplet settings is an interesting direction. 
Finally, the observation of density ripples during bubble destabilization motivates the exploration of the dynamical generation and properties of dispersive shock waves in the 2D droplet settings.
Here, it would be important to realize for instance oblique~\cite{hoefer2017oblique} dispersive shocks which are known to occur in the
absence of quantum fluctuations. Such studies are
presently under consideration and will be 
reported in future publications.

\begin{acknowledgments}
 This material is based upon work supported by the U.S. National Science Foundation under the awards PHY-2110030 and DMS-2204702 (PGK).
\end{acknowledgments}

\bibliographystyle{apsrev4-1}
\bibliography{reference}	

\end{document}